# Divination by Prompt: LLM-Mediated *Xuanxue* on Chinese Social Media

Chu'ang Li[1*], Lixuan Wang[1*], Yuqi Chen[2] & Ze Hong[1]

[1] Department of Sociology, University of Macau, Taipa, Macau SAR, China

[2] Institute for the Humanities and Social Sciences, University of Hong Kong, Hong Kong SAR, China

[*] These authors contributed equally to the manuscript

To whom correspondence should be addressed: Ze Hong (zehong@um.edu.mo)

## Abstract

The rapid proliferation of large language models (LLMs) has produced a striking cultural practice: using conversational AI for divination. This paper offers one of the first systematic studies of LLM-mediated divination in the context of *Xuanxue* (玄学), an internet-native umbrella term for mystical and spiritual practices on Chinese social media. Using a mixed-methods design, we analyze 23,000+ posts and comments from Xiaohongshu and conduct 32 semi-structured interviews with users and professional diviners. Users primarily consult LLMs about pragmatic concerns—romantic relationships, careers, exams, and in-game gacha draws—via two intersecting pathways: trend-driven curiosity enabled by viral visibility and zero-cost access, and event-driven anxiety under conditions of uncertainty. A defining feature is collaborative prompt refinement, which turns users into active prompt engineers. Among commenters expressing a clear stance, perceived efficacy skews positive, with "accuracy" often justified through biographical fit and retrospective confirmation, consistent with Barnum and confirmation bias. Users also develop verification practices such as repeated trials and cross-model comparison. Professional diviners, by contrast, portray LLMs as lacking the "spiritual power" required for genuine divination, reflecting both ontological commitments and economic boundary-work. We also show how participants navigate tensions between scientific and metaphysical frames when interpreting AI-generated readings. Situating these findings in anthropological and cognitive-evolutionary theories of divination, we argue that LLM divination preserves core functions of traditional practice while introducing scalability, repeatability, and prompt-driven co-production that reshape how divinatory authority is constructed and evaluated.

# 1. Introduction

The recent proliferation of large language models (LLMs) has begun to reshape the informational and communicative landscapes of modern society, permeating nearly every aspect of human life (Clusmann et al., 2023; Geng et al., 2024). In simple terms, LLMs are a form of artificial intelligence trained on vast datasets of text and code, enabling them to generate sophisticated, human-like text in response to user prompts (Shanahan, 2024). This technological advancement has had a profound societal impact, altering modes of communication and economic productivity, as well as how individuals search for and interact with information (Dvorak et al., 2025; Eloundou et al., 2024). However, the rise of LLMs is not without its complexities and risks, including concerns about the propagation of misinformation (Garry et al., 2024), misalignment of human-machine priorities (Kirk et al., 2024), and the potential for algorithmic biases to marginalize minority viewpoints (Navigli et al., 2023).

Amidst this widespread integration, one conspicuous and rapidly growing application of LLMs is their use for divination. This phenomenon, however, is not entirely novel but rather the latest evolution in a longer history of digital divination (Davies, 2024; Ruah-Midbar, 2014). Since the early days of the internet, digital platforms have offered virtual forms of traditional practices, from online astrology charts and tarot readings to mobile applications for various forms of fortune-telling (Wilson, 2024). The term "algorithmic divination" is frequently used to describe these practices (Carlson et al., 2023; St. Lawrence, 2024), as they employ computational procedures to process user inputs and generate seemingly prophetic or insightful outputs, thereby replacing traditional divinatory mechanics (which often require the interpretation of human agents) with automated, rule-based systems. The advent of LLMs, however, marks a significant methodological breakthrough, offering unprecedented flexibility and ease of use. Unlike pre-programmed apps, LLMs allow users to craft bespoke divinatory prompts, engage in fluid dialogue with the "oracle," and receive highly personalized and elaborate responses, effectively creating a highly interactive form of divinatory experience (Cheng, 2025).

The enduring appeal of divination, both ancient and digital, has long been a subject of anthropological and psychological inquiry (Hong, 2025a). As a widespread cultural practice, divination serves as a mechanism for resolving personal and social quandaries. Early anthropological theories typically approached divination as an instrumental and epistemic technology, a sincere (if flawed) attempt to generate accurate information to explain, predict, and control worldly events, akin to a form of proto-science (Frazer, 1890; Tylor, 1871). However, many later scholars, seeking to avoid the conclusion that practitioners were simply mistaken, proposed non-instrumental functions. These functionalist accounts posit that divination serves as a vital cultural tool for resolving social disputes (Burkert, 1985), legitimizing political authority (Flad, 2008), and reinforcing social norms and structure (Park, 1963; V. W. Turner, 1968). From a psychological standpoint, its practice has been interpreted as a coping mechanism to alleviate the anxiety associated with uncertainty, particularly in high-stakes domains where empirical control is limited (Malinowski, 1925). More recent cognitive and cultural-evolutionary approaches, on the other hand, argue that such social and psychological functions are only effective because participants grant the practice genuine epistemic value (Hong & Henrich, 2021). This perceived credibility can be bolstered by features such as "ostensive detachment" where the appearance of impartiality enhances persuasiveness (Boyer, 2020) or by transmission biases that favor the recall of successful divinatory episodes (Hong, 2022; Hong & Zinin, 2023).

The emergence of LLM divination provides a particularly productive arena for examining these theories. LLMs may be perceived as objective and authoritative by virtue of their technological sophistication—a quality that may function analogously to the "ostensive detachment" of traditional oracles—yet their outputs are actively shaped by users through iterative prompting, blurring the boundary between diviner and client and raising new questions about how divinatory authority is constructed and evaluated. At the same time, LLM divination offers a unique window into how people interact with conversational AI in everyday settings. In contrast to earlier digital fortune-telling tools that largely delivered fixed, menu-driven outputs, LLMs invite iterative, multi-turn interaction in which users actively steer the "reading" through prompt templates and role instructions that undergo successive refinement, effectively co-producing an oracle through dialogue in an efficient and cost-free manner. This makes visible several broader dynamics of human–AI use: how credibility and authority are established for opaque systems; how interface features such as regeneration, chat history, and follow-up questions encourage new verification strategies (e.g., repeated trials and cross-model comparison); and how communities develop and circulate practical know-how through prompt sharing and collective troubleshooting. As opaque AI systems are increasingly consulted for consequential decisions across many domains, LLM divination thus provides an especially useful case for examining how people use generative AI in everyday life.

The present paper provides one of the first systematic investigations of the emergent phenomenon of LLM divination within the specific cultural context of "*Xuanxue*" (玄学) on Chinese social media. Originally a form of Neo-Taoism, *Xuanxue* has evolved in modern internet parlance into an umbrella term for a range of mystical and spiritual practices, including divination, astrology, and folk beliefs, which have seen a major revival among Chinese youth as a source of comfort and social connection in a high-pressure society (Wei, 2025; Yimin, 2024). Through a mixed-methods approach that combines a large-scale analysis of over 20,000 social media posts and comments from the platform Xiaohongshu with 32 in-depth interviews of both human diviners and users of LLM divination, we systematically examine the motivations and perceived efficacy of this practice. Situating our findings within both classic anthropological and modern cognitive and cultural-evolutionary perspectives, we argue that LLM divination retains the core epistemic and social functions of traditional divination while fundamentally transforming its social organization: the affordances of zero cost and easy repeatability, combined with prompt-driven customization shift divinatory authority from specialized practitioners to networked communities of users, creating new dynamics of collaborative co-production alongside new vulnerabilities to well-documented psychological biases.

The remainder of this paper is structured as follows: Section 2 provides a broad overview of *Xuanxue* and algorithmic divination in the Chinese context, with a focus on Large Language Models. Section 3 details our data collection and analytical methodology. Section 4 presents our findings on the uses and perceptions of LLM divination. Section 5 discusses the theoretical implications of these findings as well as the individual and societal impact of LLM divination, and Section 6 concludes with a summary of our contributions and directions for future research.

# 2. Background and literature review

In this section, we examine the historical and contemporary state of Chinese *Xuanxue*, tracing its trajectory from traditional cosmological frameworks to the development of algorithmic divination in contemporary China, where generative AI and Large Language Models are reconfiguring ancient traditions for the modern user.

## 2.1. The past and present of Chinese *Xuanxue* cosmology and practices

In traditional Chinese thought, *Xuanxue* was a metaphysical school associated with the Wei and Jin dynasties, grounded in correlative concepts such as *qi* (vital energy), *yin-yang* (dynamic polarity), and the Five Phases, which together formed a systematic framework for understanding changes in nature, society, and the human body (Graham, 2015; Lu, 2022; Sivin, 1995). This cosmological foundation supported a wide range of prognostic practices that endured for millennia, including *Yijing* (I Ching) divination, bagua-based *feng shui*, temple lot-drawing, and natal astrology methods such as *Bazi* and *Ziwei doushu* (Bruun, 2003; Field, 2008).

In the early twentieth century, following the collapse of the imperial order and amid Republican iconoclasm and later socialist campaigns, these cosmologies and practices were increasingly branded “feudal superstition,” criticized as obstacles to scientific modernization and subjected to regulation and suppression (Goossaert & Palmer, 2011; Hong & Chen, 2026; C. K. Yang, 1970). Despite these institutional constraints, the traditions survived in academic scholarship, popular almanacs, and household rituals, while simultaneously persisting in overseas Chinese communities (Crissman, 1967). Following the Reform and Opening-up period in the 1980s, texts and expertise from Taiwan and Hong Kong began to circulate back into mainland China, sparking a revival (Smith, 2021). In recent years, certain forms of divination have even been partially re-legitimized as a form of traditional culture within broader heritage projects (G. Li, 2018).

Emerging from this complex history, contemporary Chinese *Xuanxue* is defined by two main trends. The first is the resurgence of traditional esoteric practices, and the second is the coexistence and interaction between Chinese divination and Western systems. Since the economic reforms, fortune-telling based on the *Yijing*, along with physiognomy and palmistry, has returned to both urban and rural areas (Bruun, 2003). As the socioeconomic landscape changed, practitioners moved into public spaces to commercialize their work. Many began operating as street diviners or opened professional studios for fortune-telling and *feng shui* (Homola, 2018; Ren, 1993). At the same time, the importation of Western divinatory forms such as Tarot and astrology has provided alternative frameworks for coping with life's uncertainties (Fu et al., 2023; Zhang & Wang, 2025; Zhuang, 2025), and in some cases encouraged cross-cultural translation and fusion (Singh, 2024).

## 2.2. The algorithmic turn in Chinese *Xuanxue*

In contemporary China, the term *Xuanxue* has drifted far from its classical philosophical roots to become a broad, internet-native label for mystical and spiritual interests (Wei, 2025). While the term encompasses a wide variety of phenomena and techniques that seemingly defy scientific explanation such as *feng shui* and talismanic practice, its primary practical application is in various forms of divination. These topics circulate widely across China's major social media platforms such as Weibo (equivalent to X/Twitter), Bilibili (equivalent to YouTube), and Xiaohongshu (equivalent to Instagram) (C. Liu, 2025), where traditional divination is increasingly mediated through digital formats.

The digitization of these practices is not unprecedented. As early as the 2000s, websites offering *Yijing* fortune-telling emerged across Taiwan, Hong Kong, and mainland China, attracting large numbers of users looking for personalized readings (Kuo, 2009). However, the rise of social media has allowed algorithmic fortune-telling to expand into new formats. Fu et al. (2023), for instance, examined interactive Tarot videos on Bilibili. They found that creators were not simply copying Western tarot traditions but actively adapting them for a local audience. By incorporating Chinese cultural symbols and encouraging viewers to discuss their results in the comments, these videos have transformed a foreign practice into a familiar source of comfort. This approach has proven particularly effective for young Chinese women, who make up a significant portion of the audience and often turn to online readings for emotional support and self-reflection (Zhuang, 2025).

A more recent development is the use of generative AI and large language models (LLMs) as tools for divination, a practice that has rapidly become a recognizable pattern of cultural consumption in China's digital public sphere (Wang, 2025). On the one hand, designers in human–computer interaction and software developers leverage the dialogic, flexible interactivity of AI to produce anthropomorphic, "consultation-like" experiences (Lai & Zhou, 2023). A prime example is *Cece*, a mobile app popular among Chinese youth, boasting over 40 million registered users and 2.6 million monthly active users (Xinyan Group, 2024). *Cece*'s algorithms combine user data such as birthdays, zodiac signs, and *Bazi* to generate daily and monthly fortune forecasts alongside actionable guidance. Recently, the app introduced a paid "AI Virtual Counselor." Although framed as mental health support, this feature effectively functions as an intelligent, always-on fortune teller[1]. Similar services are proliferating across the digital ecosystem, such as *Wen Zhen Bazi*, an online *Bazi* charting platform, and *Bu Meng Dao*, which offers AI tarot, zodiac charts, and dream interpretation. Furthermore, numerous other generative AI-based fortune-telling models are currently under development (Ha, 2025; X. Li, 2025).

On the other hand, general-purpose LLMs are themselves being appropriated for fortune-telling in everyday practice, particularly following their recent commercialization and widespread availability (Ha, 2025; Wang, 2025). Although empirical research remains sparse, early reports suggest that people (especially the young) are turning to these tools as accessible, low-cost alternatives to traditional divination. Users utilize them not only for predictive insights regarding careers and relationships but also as a source of non-judgmental psychological support to help cope with existential uncertainty and anxiety (Hu, 2025). There is also evidence suggesting that some users treat models like ChatGPT and DeepSeek as pliable oracles, steering their outputs through iterative prompting. For instance, Ling (2025) documents how users engage in "re-telling" practices. Instead of accepting the AI's first answer, some users will keep adding context and tweaking their prompts until the model yields a desirable response, essentially validating their pre-existing beliefs or emotional states. Other reports suggest that systematic prompt refinement is viewed by users as a way to achieve more "accurate" results over time (F. Yang, 2025), indicating the adoption of a pragmatic attitude similar to that found in traditional divination practices (Hong, 2025a).

# 3. Methods and data description

In the present study, we employed a mixed-methods approach to investigate the emerging phenomenon of using large language models for *Xuanxue* purposes. Specifically, we scraped data from the Chinese social media platform Xiaohongshu (often described as China's equivalent of Instagram) with over 300 million monthly active users and more than 100 million content creators, 70% of whom are female and 85% born after 1995 (Qian, 2025). We selected this platform because it represents a unique intersection of technological and cultural trends: it is both a hub for the rapid proliferation of AI-generated content (Mu, 2024) and home to a user base deeply engaged in ritual and spiritual practices (Wei, 2025). Complementing the social media analysis, we conducted in-depth semi-structured interviews with both human diviners (who had some familiarity with LLM divination) and users of LLM divination. This qualitative component is essential for triangulation, allowing us to uncover the subjective motivations and interpretive nuances that cannot be fully captured through platform data alone

[1] See https://ai-bot.cn/cecelive/; accessed Jan 22, 2026

## 3.1. Social media data

### 3.1.1 Data Scraping

We collected data using a custom Python script targeting the Xiaohongshu web interface. The search query focused on the keyword "DeepSeek 玄学" (DeepSeek *Xuanxue*). This specific phrasing was chosen to capture the intersection of two trends: the widespread adoption of DeepSeek, a domestic LLM that surged in popularity in early 2025 due to its accessibility without a VPN, and the use of *Xuanxue* as a coded term for divination. Using *Xuanxue* avoids platform censorship mechanisms, which often flag explicit terms like "Fortune Telling" as superstition (Fu et al., 2023).

For each identified post ID, the script accessed the platform's detail API to retrieve comprehensive metadata. This included basic attributes (e.g., title, body text, hashtags, publication timestamp, and IP location), multimedia links, and interaction metrics (e.g., likes, shares, saves, and comments). To capture the full nested comment structure, we employed a recursive retrieval strategy: first-level comments were retrieved first; upon detecting second-level replies, the script iteratively queried the corresponding API to capture all subsequent responses. The resulting dataset included all posts and associated comments published under the target keyword up to April 10, 2025. The final corpus consisted of 232 posts and 23,340 comments, comprising 11,200 first-level comments and 12,140 corresponding replies. To improve data quality, promotional content was filtered out: only standard user-generated posts (classified as "note" in the platform's metadata) were retained, while advertisements (identified by the platform's metadata labels) were excluded.

### 3.1.2 Content analysis via LLM

To analyze the comments from social media data, we employed a hybrid approach combining Large Language Model automation with human validation. DeepSeek-V3.1 served as the automated coder, selected for its superior performance in Chinese natural language understanding and complex logical reasoning (A. Liu et al., 2024); in our exploratory analysis, it generally outperformed ChatGPT 4.0 in terms of its consistency with human annotation. To optimize model performance, we developed structured prompts based on the CO-STAR framework (Context, Objective, Style, Tone, Audience, Response) (Ohalete et al., 2025). These prompts not only detailed the coding criteria but also instructed the model to incorporate contextual information (i.e., original post and first-level comments) to accurately interpret specific internet slang, such as "*dun* 蹲" (literally "squat," meaning to mark the post to await updates) and "*jie yunqi* 接运气" (literally "catch luck," meaning to claim similar good fortune).

We selected three coding dimensions because they capture complementary aspects of the phenomenon central to our research questions: what people consult LLM "oracles" about (Topic Classification), how they evaluate and negotiate credibility (Attitudinal Orientation), and how users participate in and transmit prompting as a practical technique (Prompt Exchange). Specifically, our coding scheme (full prompts are provided in the Supplemental Information) included:

> *Topic Classification* identifies the user's primary focus and usage context. We defined 11 mutually exclusive categories (e.g., Romantic Relationships, Job/Study, Wealth, and Health). To resolve ambiguities in multi-topic comments, seven hierarchical tie-breaking rules and five supplementary guidelines were established.
>
> *Attitudinal Orientation* measures the user's inferred belief in the accuracy of LLM divination using a 5-point scale ranging from -2 (strong disbelief) to 2 (strong belief), with 0 indicating a neutral or ambiguous stance.
>
> *Prompt Exchange* serves as a binary variable (0/1) indicating whether the comment involves requesting, sharing, or discussing specific prompts, capturing the depth of technical interaction in Xiaohongshu communities.

To assess the reliability of this approach, we conducted an inter-rater reliability test comparing DeepSeek-V3.1 with human coding. A random sample of 224 comments was independently coded by two trained human coders; disagreements were resolved through discussion to produce a consensus "ground truth." The same sample was then coded by DeepSeek-V3.1. For the two human coders, agreement was evaluated using Cohen's Kappa; Fleiss's Kappa was then used for the three-way comparison that included the LLM. The two human coders achieved high inter-rater agreement across all three dimensions (Cohen's Kappa > 0.90), and DeepSeek-V3.1 achieved moderate to substantial agreement with the human consensus (overall Fleiss's Kappa > 0.60) (see Supplemental Information for details). DeepSeek-V3.1 was then used to code the full dataset. We note that the use of a DeepSeek model as both the analytical tool and the object of study is coincidental; the coding model (V3.1) differs from the consumer-facing versions employed for divination and was selected solely on the basis of its benchmark performance in Chinese-language tasks.

### 3.1.3. Additional data collection

To ensure the study captures the evolving dynamics of the "DeepSeek *Xuanxue*" phenomenon and to address the temporal gap following the initial automated scraping, a supplementary phase of data collection was conducted. Covering the period from April 11, 2025, to December 23, 2025, this phase employed a purposive sampling strategy rather than exhaustive scraping. Using the same keyword search, we manually curated 25 representative posts, together with 192 corresponding comments. The selection criteria prioritized high user engagement (i.e., top-ranking posts in search results) and thematic diversity, ensuring that these posts reflected the prevailing community sentiments and emerging practices during the latter half of 2025. This smaller, high-quality dataset serves as a qualitative longitudinal check to validate whether the patterns identified in the primary corpus (Section 3.1.1) persisted or shifted over time.

## 3.2. Interview data collection

### 3.2.1. User participants

User participants were recruited through purposive sampling on Xiaohongshu and through snowball sampling within university student networks in mainland China. We selected 26 respondents who met at least one of two inclusion criteria: consistent use of DeepSeek for divination over the three months preceding the interview, or some experience with the practice combined with active participation in related discussions on Xiaohongshu between January 2025 and April 2025. We conducted semi-structured interviews lasting 30–45 minutes via online video call or by telephone. The interview guide covered four areas: (1) usage habits (adoption timeline, frequency, and typical inquiry scenarios); (2) perceived effectiveness (subjective accuracy of DeepSeek outputs and the perceived role of prompting strategies); (3) comparative perspectives (how respondents position LLM divination relative to human fortune-telling and other digital divination apps); and (4) engagement and outlook (willingness to pay, sharing behaviors, and whether they would recommend the practice to friends). Questions were primarily open-ended to allow unexpected themes to emerge.

### 3.2.2. Practitioner (diviner) participants

Concurrently, we recruited 6 practitioners specializing in *Xuanxue* via Xiaohongshu. In this study, practitioners are defined as individuals possessing specialized proficiency in divinatory arts, such as Tarot, *Bazi*, or hexagram interpretation. These individuals provide paid consultative services to others, operating either on a freelance or full-time professional basis, and have had some exposure to LLM-based divination tools like DeepSeek. Semi-structured interviews with this group addressed: (1) routine professional practices and pricing; (2) attitudes toward LLM divination; (3) the integration of LLMs into their existing workflows; and (4) the professional impact of LLM divination on current human-to-human divination.

### 3.2.3. Interview data processing

All 32 interviews were audio-recorded with consent, transcribed using automated tools followed by manual verification, and translated into English for further analysis. Participants received a small incentive of 10-15 RMB (roughly 1.5-2 USD) for their time. Participant demographics are detailed in Table 1, and the complete interview protocols are available in the Supplemental Information.

| Participant | Age | Gender | Occupation | Identity |
|---|---|---|---|---|
| P1 | 30 | Female | NA | Practitioner (Full-time) |
| P2 | 25 | Female | Undergraduate student | Practitioner (Part-time) |
| P3 | 26 | Female | Graduate student | Practitioner (Part-time) |
| P4 | 25 | Female | NA | Practitioner (Full-time) |
| P5 | 28 | Male | NA | Practitioner (Full-time) |
| P6 | 24 | Female | Graduate student | Practitioner (Part-time) |
| U1 | 24 | Female | Graduate student | User |
| U2 | 23 | Female | Undergraduate student | User |
| U3 | 24 | Female | Lawyer | User |
| U4 | 25 | Male | NA | User |
| U5 | 27 | Female | Internet industry | User |
| U6 | 25 | Female | NA | User |
| U7 | 22 | Female | Graduate student | User |
| U8 | 22 | Female | Graduate student | User |
| U9 | 22 | Male | Undergraduate student | User |
| U10 | 22 | Female | Graduate student | User |
| U11 | 22 | Female | Undergraduate student | User |
| U12 | 21 | Female | Undergraduate student | User |
| U13 | 22 | Female | Graduate student | User |
| U14 | 22 | Male | Undergraduate student | User |
| U15 | 29 | Male | Operation manager | User |
| U16 | 23 | Female | E-commerce customer service | User |
| U17 | 26 | Female | Administration staff | User |
| U18 | 24 | Female | New media operation manager | User |
| U19 | 27 | Female | Online business owner | User |
| U20 | 29 | Male | Asset operation manager | User |
| U21 | 28 | Male | Customer service manager | User |
| U22 | 19 | Female | Undergraduate student | User |
| U23 | 25 | Male | Graduate student | User |
| U24 | 21 | Female | Undergraduate student | User |
| U25 | 22 | Female | Graduate student | User |
| U26 | 22 | Female | Graduate student | User |

*Table 1. Demographic information of interview participants.*

# 4. Results

## 4.1. Thematic analysis of DeepSeek *Xuanxue*: what do users ask?

Our analysis of 23,340 comments and corresponding interview data reveals that DeepSeek is rarely employed for abstract philosophical or spiritual inquiry. Instead, users mobilize the LLM primarily as a pragmatic instrument for managing the anxieties inherent in modern life. Table 2 provides a descriptive taxonomy of the topics found in the dataset, accompanied by illustrative examples to clarify the nature of these inquiries.

| Topics | Description | Example posts/comments (translated into English) |
|---|---|---|
| Numerology | Using DeepSeek/LLMs to generate overall fortune readings via systems such as the *Yijing*, *Bazi* (Eight Pillars), and astrology. | "DeepSeek's cyber fortune-telling is even more accurate than Taoist priests."<br>"DeepSeek *Xuanxue* prompt: an ultra-detailed life reading." |
| Romantic Relationships/Marriages | Using DeepSeek/LLMs to predict romantic outcomes—e.g., when one might meet a partner, whether reconciliation is possible, or whether to break up, marry, etc. | "DeepSeek hit me with a soul-crushing truth: the dating reality 99% don't dare face."<br>"DeepSeek says my boyfriend will contact me." |
| Job/Study/Exam | Using DeepSeek/LLMs to predict employment outcomes, exam results, admissions, certifications, and related performance questions. | "AI fortune-telling god—I got hired!!"<br>"Let's use DeepSeek divination to see if I can pass the driving test."<br>"Teacher certification results are coming—who used DeepSeek to estimate their score?" |
| Gaming/Gacha | Using DeepSeek/LLMs to predict gaming outcomes, especially gacha draws (probability-based character/item pulls). | "DeepSeek divination: ten-pull hit!"<br>"Wait—did DeepSeek's 'geolocation divination' really help me find the 'heavenly treasure stone'?" |
| Health/Pregnancy/Period | Using DeepSeek/LLMs for divinatory readings about health conditions, | "Using DeepSeek to cast a hexagram to predict pregnancy." |
| | pregnancy timing, and menstrual cycles. | “DeepSeek predicts my period.” |
| Human Diviners Advertising | Comments posted by offline/online fortune-tellers under DeepSeek *Xuanxue* content to attract customers to their own paid services. | “I’ve worked in this field for 20+ years… I’ll tell you the good and the bad, how to avoid pitfalls and make up deficiencies… only a few dozen yuan… Truly capable masters aren’t that expensive—don’t get scammed.” |
| Money/Wealth/Lottery | Using DeepSeek/LLMs to predict financial luck, wealth prospects, lottery outcomes, winning numbers, and related questions. | “DeepSeek’s get-rich secrets: the lottery guide is here!!”<br>“DeepSeek predicts lottery numbers—get rich or lose big?”<br>“What about the child’s wealth?” |
| Pursuing Celebrities/Idols | Using DeepSeek/LLMs to assess “compatibility” with idols, predict concert ticket success, seating/locations, and related fandom outcomes. | “DeepSeek’s concert seating prediction is really amazing.”<br>“You’re a top *Bazi*/fortune-telling master, analyze whether these idols suit me and whether they bring me luck.” |
| Pets | Using DeepSeek/LLMs to divine the afterlife status of deceased pets (e.g., reincarnation), and occasionally to locate missing pets. | “I used DeepSeek to calculate my karmic bond with my lost dog.”<br>“I asked DeepSeek, via mysticism, about my deceased puppy.” |
| Things Lost and Found | Using DeepSeek/LLMs to locate lost objects and infer their whereabouts. | “DeepSeek, you’re insanely accurate at finding things.”<br>“DeepSeek found the earphones I’d lost for three days!” |

*Table 2. Topic classifications of posts/comments in the social media dataset, with descriptions and illustrative examples.*

As illustrated in Figure 1, these topics are not distributed evenly. Most notably, the high frequency of general “Numerology” (n = 6,307) suggests a broad interest in the mechanics of fortune-telling with LLMs, particularly the use of DeepSeek for *Yijing*–style divination and personalized fortune readings. Typical posts describe users entering highly structured prompts (see Section 4.2 for a detailed discussion of prompt sharing) that instruct DeepSeek to perform as an expert practitioner, for example:

> Input: “You are a top-tier *Bazi* (eight pillars) master. Please analyze my *Bazi*: Year pillar: xx; Month pillar: xx; Day pillar: xx; Hour pillar: xx. Gender: x. Place of birth: xxx. Current residence: xxx. Strictly follow *Bazi* theory and methodology. First, examine my chart and major cycles. Then verify my life events in detail. Thank you.”

In this instance, the LLM is asked to assume the role of a traditional diviner and “strictly follow *Bazi* theory and methodology,” implying that it is the user’s belief in the validity of the *Bazi* system itself that lends plausibility to the verdict generated by DeepSeek.

Although most posts in this category focus on traditional Chinese numerology, there are also mentions of Western approaches, such as tarot cards and astrology. Occasionally, users synthesize these different traditions into a single prompt to “cover all the bases.” One such example reads:

> You are a top-tier global fortune teller. Based on my *Bazi*: “Year xx, Gregorian month xx, day xx, hour xx; Lunar month xx, day xx; Birthplace: xx Province, xx City; Gender: Female”, and combining Tarot, Astrology, the *Yijing*, and the *Bagua* (Eight Trigrams), describe how many past lives I have had. Provide a profile for each, including appearance, height, historical period, family conditions, educational background, personality, age, and marriage. Additionally, provide physical evidence of these past lives present on my current body.

This prompt exemplifies a creative attitude where users try to ascribe an epistemic authority role to the LLM by enumerating a number of divination traditions. As will be discussed in more detail later, such an attitude highlights the pragmatic mindset of users **who** treat the LLM as a general-purpose, low-cost oracle whose authority can be amplified by stacking traditions, rather than as an expert system anchored in any single coherent metaphysical framework.

The second-largest category, “Other” (n = 5,763), consists mainly of short, non-substantive comments, but also captures extended, topic-specific exchanges that are not directly about DeepSeek’s metaphysical functions. For instance, beneath posts regarding romantic readings, commenters often shift into wide-ranging philosophical debates about the nature of intimacy, suggesting that posts framed around DeepSeek’s *Xuanxue* uses can function as conversation starters, prompting sustained discussion that extends well beyond the divinatory output.

Among the remaining categories, the most common themes are Romantic Relationships/Marriage (n = 3,411) and Job/Study/Exams (n = 3,195), followed by Gaming/Gacha (n = 1,470) and Health/Pregnancy/Period (n = 910). These frequencies map onto the primary stressors of contemporary Chinese youth (Jankowiak & Moore, 2012).

In the domain of romantic relationships, users frequently turn to DeepSeek to infer the unobservable interiority of others—asking the LLM to predict whether an ex-partner will reach out, or to assess compatibility with a crush. The comments reveal a desire to make the ambiguity of modern dating interpretable through algorithmic means. Similarly, the Job/Study category reflects the intense competitive pressures facing Chinese youth. Here, the AI is consulted not just for prediction (e.g., *"Will I pass the postgraduate entrance exam?"*), but for strategic optimization, with users asking the model to calculate the most auspicious times to submit applications or to analyze the "fate" of a specific career path. Additionally, the Health/Pregnancy/period category reflects the demographics of Xiaohongshu, which has a predominantly female user base.

Across these high-frequency topics, a shared undercurrent is uncertainty coupled with limited personal control. Users turn to DeepSeek to infer another person's feelings, estimate exam chances, or strategize around probabilistic outcomes. This logic is especially visible in the Gaming/Gacha category, where players pay for randomized "draws" to obtain rare characters or items. Faced with these odds, users recruit DeepSeek as a divinatory aid to identify "optimal" draw times. A typical prompt template reads:

> You are a master of auspicious timing. I will be pulling from the *Tian Shang* prize pool in *Niushuihan* (game name) after [date and time]. The *Tian Shang* Stone is guaranteed within [number] pulls, and I currently have [number] pulls remaining. Please calculate an exact minute to ensure I obtain the *Tian Shang* Stone or *Xiangrui* within these [number] pulls. I was born on [date]. Thank you.

When original posts offer explicit prompts like the one above, the comment sections often evolve into feedback loops, where other users report whether the prompt generated accurate results for them.

Finally, the category of Human Diviners Advertising reveals a professional tension. Many self-identified fortune-tellers promote paid services by explicitly questioning the efficacy of DeepSeek. They often bolster their credibility by claiming identities such as "Daoist priest" or asserting ties to recognized clergy. For example, one user commented:

> DS (DeepSeek) is totally inaccurate... I tested it myself and it was complete nonsense. Later, I was lucky enough to consult a master who was accurate. AI really falls short in this area.

Such remarks appear designed not only to discredit DeepSeek, but to prompt curiosity about the unnamed "master," creating a pathway for customer acquisition. This rhetorical script recurs across the dataset, with another user claiming: *"DS never got me right... I got so anxious that I consulted a master who gave detailed insights... It all came true."* In both cases, DeepSeek is portrayed as inaccurate and generic, while the human diviner is framed as precise and timely, with outcomes validated by reality. A detailed discussion of the perceived efficacy of DeepSeek divination is presented in Section 4.4.

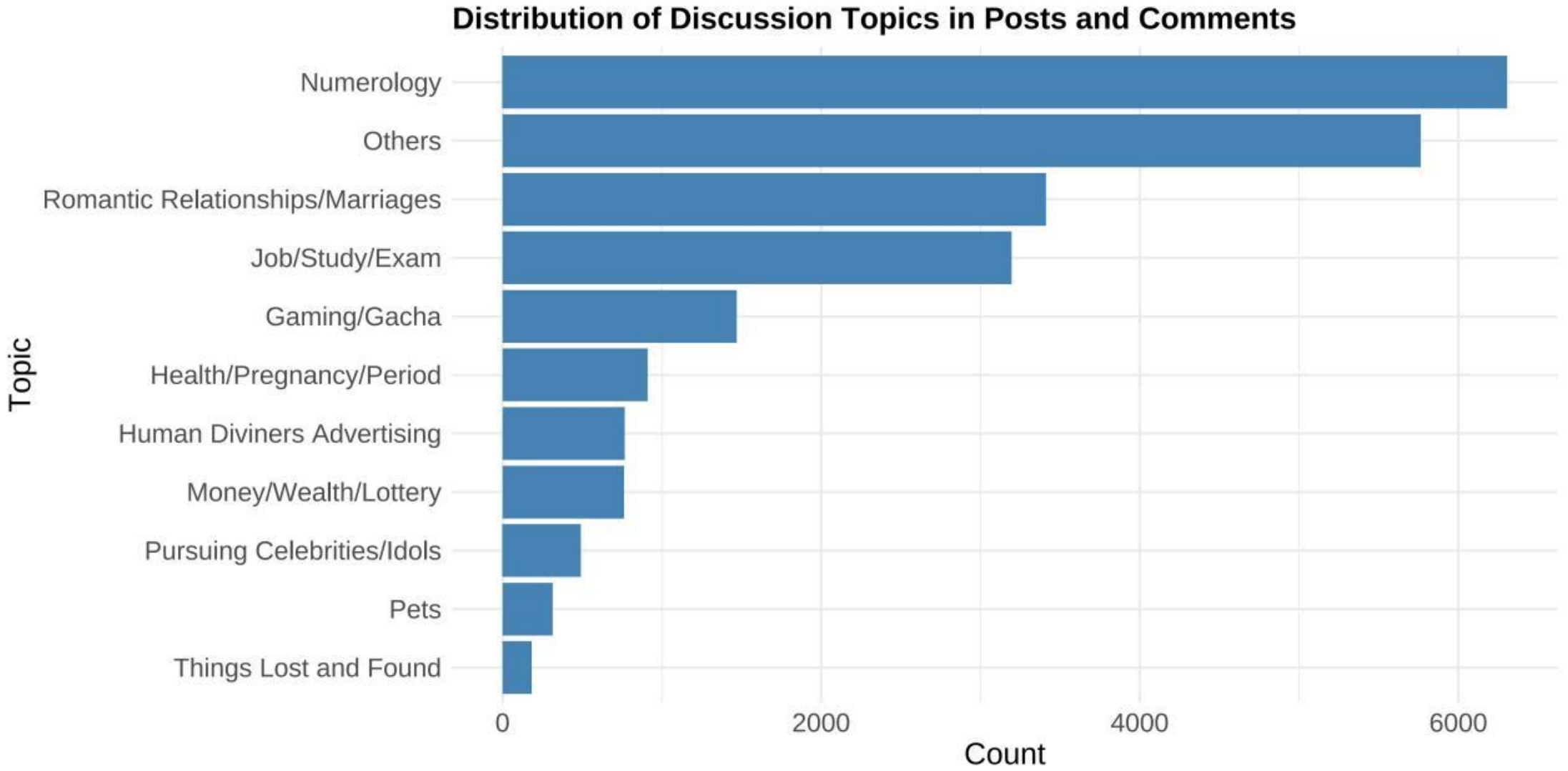

*Figure 1.Distribution of topics in Xiaohongshu posts/comments. Bars represent the frequency of coded categories across the dataset.*

## 4.2. Divinatory prompt exchange and sharing

A defining characteristic of the DeepSeek *Xuanxue* community is the extensive exchange and collaborative refinement of input strategies. This behavior reflects an implicit assumption that divination accuracy in LLMs is largely contingent upon the structural quality of the prompt (Knoth et al., 2024; Sahin, 2024). This prompt-engineering logic transforms users from passive observers into active practitioners, revealing a highly pragmatic attitude in wanting to replicate the divinatory procedure themselves. Our analysis of comment sections reveals a pervasive culture of technical solicitation, driven by the acute awareness that varying prompts yield rather different results. As one user observed, "*There are so many templates on Xiaohongshu; I tried them separately and found that each one calculates a different result.*" Consequently, users frequently post specific, vernacular requests such as "*Begging for the prompt, baby,*" "*How exactly did you ask?,*" and "*What is the input template for this?*" In response, successful users frequently act as open-source contributors, generously pasting their exact prompt structures to help peers debug their own interactions with the LLM.

Figure 2 quantifies this phenomenon, displaying the proportion of posts and comments explicitly dedicated to prompt exchange across different topics. The data reveal that prompt-sharing is ubiquitous, appearing in 10–25% of interactions across most categories. This high prevalence underscores that *Xuanxue* on Xiaohongshu is very much about trying and potentially mastering a technical skill.

The collaborative nature of these exchanges often mirrors software debugging or iterative design. For example, in discussions regarding lottery predictions, users engaged in extended workshop-like exchanges on prompt formulation. When one user expressed frustration with vague results, another advised: "*Just tell it [DeepSeek] what you want. If the results are off, keep correcting and refining it.*" A third user proposed a strategic pivot in logic: "*I don't think using DS should be about getting it to recommend the exact winning numbers, that's unrealistic. Instead, refine your prompt phrasing to get it to suggest combination tickets (plans for selecting lotto numbers). This would be a better strategy.*" Similar prompt-centered discussions recur beneath posts on job hunting, celebrity fandom, and gacha draws, illustrating how users treat the LLM as a malleable tool that requires skillful calibration.

However, a notable exception to this culture of sharing is found in the "Human Diviners Advertising" category, which exhibits the lowest proportion of prompt exchange. This deviation highlights a fundamental conflict in business models. While AI users treat divination as a reproducible technical skill (effectively "open-sourcing" the magic), professional fortune-tellers rely on information asymmetry to sustain their livelihood. For a human diviner, the specific method of calculation is often proprietary capital; to share the "prompt," or even to concede that a well-crafted prompt *could* generate an accurate verdict, would undermine the perceived value of paid consultations. As a result, posts in this category more often discourage reliance on LLM outputs and emphasize the need for human expertise, rather than participate in template sharing.

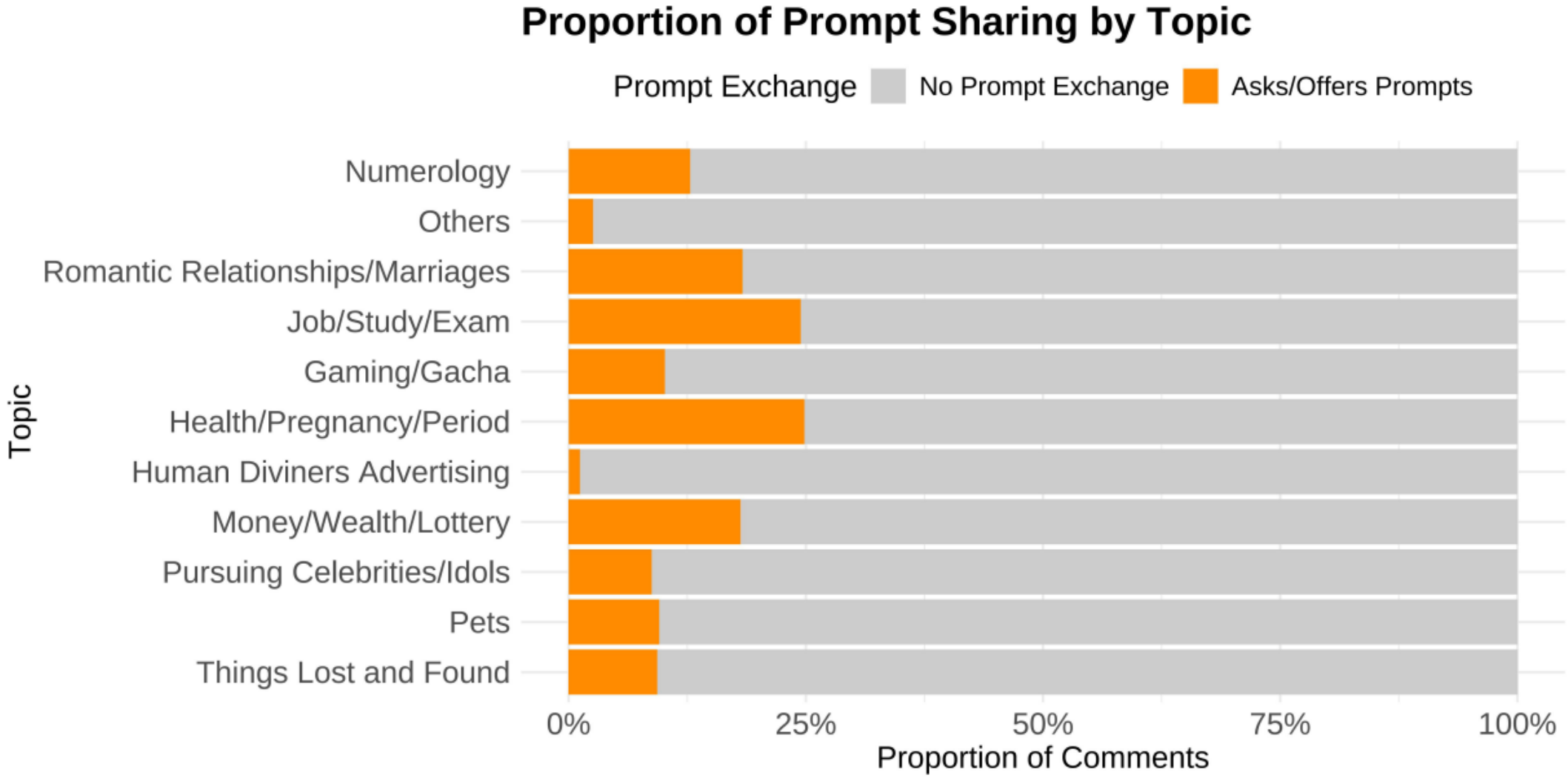

*Figure 2. Proportion of prompt sharing by topic in Xiaohongshu comments. Stacked bars indicate, for each topic category, the percentage of comments that ask for or offer prompts/templates versus comments with no prompt exchange.*

## 4.3. Dual pathways of participation in DeepSeek *Xuanxue*: curiosity and anxiety

Having examined the technical mechanics of prompt exchange, we now turn to the underlying motivations driving users toward LLM-based divination. Our interview data reveal two distinct, yet often intersecting, motivational pathways: trend-driven curiosity and event-driven anxiety.

The first pathway was propelled by the sudden popularity of DeepSeek and Xiaohongshu's viral platform mechanics. From January to May 2025, DeepSeek *Xuanxue*–related content surged on Xiaohongshu. Discussion posts appeared with increasing frequency in users' feeds, while comment threads became sites of prompt exchange, where users sought to "optimize" inputs to produce more "accurate" divination. Many interviewees reported that their initial engagement was not born of a specific spiritual need, but rather a response to this ubiquity. Circulating through prompt-sharing posts and copy-paste templates, DeepSeek *Xuanxue* functioned as a zero-cost, low-barrier entry point to divination. As one interviewee (U9) recalled: "*I first started using DeepSeek for divination around early March this year (2025). I came across it while scrolling through Xiaohongshu, many people would provide you with a prompt, and then you'd use that prompt to calculate things.*" Others emphasized the appeal of risk-free experimentation: "*After seeing some posts on Xiaohongshu, I figured it wouldn't cost me anything, so I decided to give it a try*" (U26).

This experimentation—driven by novelty and curiosity, and thus distinct from more conventional functions of divination—lends DeepSeek's *Xuanxue* divination a more entertainment-oriented character. Scholarship on religion in the new media era suggests that online spiritual practices often operate less as stable belief systems than as pop-cultural, participatory repertoires embedded in everyday media use (Campbell & Tsuria, 2021). Consistent with this view, some users approached DeepSeek less as an authoritative oracle than as a playful, generative interlocutor. As one interviewee put it: "I also ask [LLMs] silly questions, like having DeepSeek calculate what kind of weapon (*wu qi*) I'd be if I were one, based on my birth chart." (U1)

Underpinning this playful engagement is the sheer convenience of the technology, which numerous interviewees cited as the most obvious advantage of LLM divination. Additionally, DeepSeek provides a confidential space that eliminates the interpersonal pressures and social debts inherent in human consultations. Many interviewees described feeling constrained when consulting human fortune-tellers, worrying that asking trivial or excessive questions might violate social etiquette. By contrast, the AI interface removes the fear of judgment. As one interviewee (U10) noted:

> DeepSeek's greatest advantage is the lack of psychological burden. When consulting a human, you might worry whether your question makes sense, if it warrants a reading, or if the practitioner is even willing to answer it. With AI, however, there are no taboos. For instance, I would never ask a human fortune teller which celebrity I should chase, but with AI, I can ask anything.

Similarly, another interviewee (U14) highlighted the perceived privacy of LLM divination:

> If you ask a real person [diviner], you owe them a favor (*renqing*), or you have to pay them money to get a reading. With this, it saves effort and ensures confidentiality. At least I feel that, well, it won't be seen by others; it won't be uploaded to the internet for others to see.

Yet, the appeal of DeepSeek *Xuanxue* extends beyond low-stakes entertainment or casual curiosity. A second, more traditional pathway exists alongside this playfulness: situational anxiety. When faced with concrete high-stakes uncertainties such as exams, relationship turbulence, or career pivots, participants reported a heightened desire for clarity and control. In these instances, the practice aligns more closely with anthropological understandings of divination as a mechanism to manage uncertainty and offer a sense of knowledge and control in anxiety-inducing situations (V. Turner, 1975).

These two pathways often intersect. While platform visibility (the trend) creates widespread awareness of the technology's affordances, personal anxiety provides the immediate impetus to participate. One respondent (U2) illustrated this convergence vividly:

> I'm preparing for the postgraduate entrance exam and feeling quite anxious... This thing is pretty popular now. Even last week (May 2025), while taking the subway home, I overheard people discussing how to use AI for fortune-telling... I also frequently see people using DeepSeek on the subway to calculate various things. After the Spring Festival, once this trend gained momentum... some even ask others if the predictions are accurate.

Crucially, for some users following this second pathway, belief in the supernatural is secondary to the functional need for emotional relief. Some interviewees explicitly stated that they turn to LLM divination for psychological comfort during high-stress periods without actually believing in the validity of the output. As one respondent (U25) explained:

> Because actually, speaking from the bottom of my heart, really, I don't really believe in it. But perhaps during this period of preparing for the postgraduate entrance exam, it really does make me less anxious. It serves as a psychological anchor, a form of psychological comfort. It's just a crutch, but I don't believe it's real. (U25)

At the same time, it would be misleading to assume that users uniformly adopt such a detached stance. While some treat LLM divination primarily as a coping tool, others report being genuinely struck by its apparent predictive accuracy, and still others remain firmly skeptical. To capture this full range of orientations, we next examine users' assessments of the perceived validity and efficacy of DeepSeek *Xuanxue*.

## 4.4. Perceived efficacy of DeepSeek *Xuanxue*

### 4.4.1. General pattern of efficacy perception

In both Xiaohongshu posts/comments and interviews, users of DeepSeek *Xuanxue* expressed mixed attitudes regarding its efficacy. Based on the coding scheme outlined in Section 3, we have categorized each post/comment into one of the five categories (from Strongly Disbelieve to Strongly Believe; see Figure 3). Overall, neutral/unclear evaluations constitute the largest share of comments (9,302; ≈39.5%). This predominance of indeterminate positions likely reflects the structure of comment threads on Xiaohongshu. Many second-level replies shift away from assessing the "accuracy" of DeepSeek's divination and instead develop into broader, topic-centered exchanges.

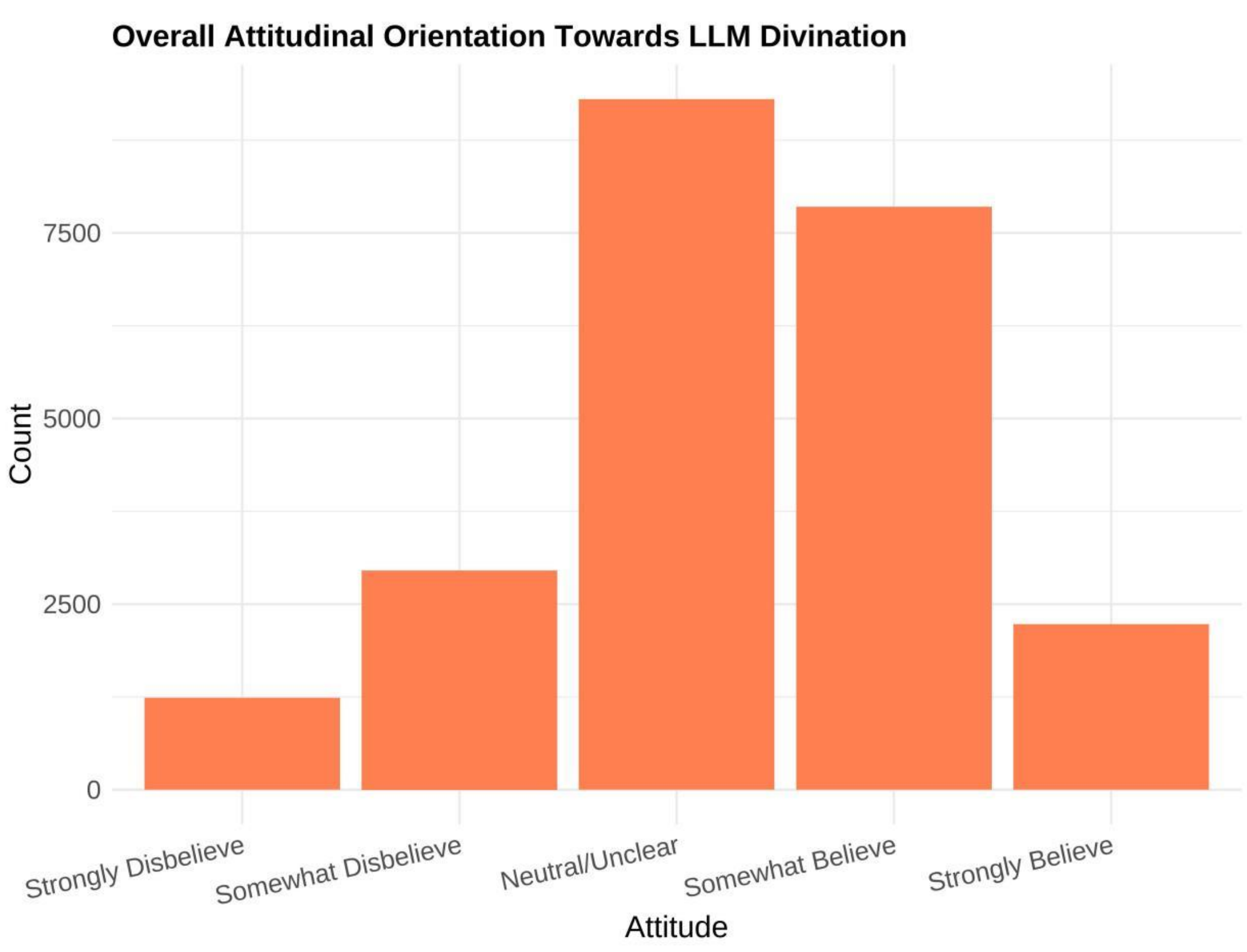

*Figure 3. Distribution of attitude toward the efficacy of LLM-based divination in Xiaohongshu comments. Counts show responses by stance category (from strongly disbelieve to strongly believe).*

Importantly, among posts and comments that express a discernible evaluation, attitudes skew positive. "Somewhat believe" (7,851; ≈33.3%) and "strongly believe" (2,229; ≈9.5%) together account for 10,080 entries (≈42.8%), whereas "somewhat disbelieve" (2,953; ≈12.5%) and "strongly disbelieve" (1,237; ≈5.2%) total 4,190 entries (≈17.8%). In other words, beyond the ambiguous/neutral portion, users are far more likely to voice belief than skepticism toward DeepSeek's *Xuanxue* practices. This pro-belief tilt is most evident in "review-like" comments that cite personal experience as evidence of accuracy. For example, one user reported: "*It's so accurate! I asked about my future career path, and it listed finance-related fields while recommending I explore social media. I'm currently majoring in financial management and serve as a student council member for new media operations at my school.*" Another wrote: "*It's so accurate, it's scary—it even predicted my dad has heart disease.*" Many comments follow this same logic, treating biographical alignment and retrospective confirmation as grounds for judging DeepSeek's divinatory outputs as credible.

In our interviews, participants sometimes shared vivid personal stories in which DeepSeek *Xuanxue* generated seemingly accurate predictions in great detail. One interviewee (U3), for example, characterized DeepSeek's tarot predictions as "incredibly accurate", drawing primarily on her boyfriend's experiences with the tool:

> My boyfriend lives abroad and drives a lot, but he's not super familiar with the local traffic rules, like pushing it at yellow lights or overstaying in parking. He keeps asking the AI (DeepSeek) things like, 'Did I get any traffic violations today? Did I run any red lights?' Each time he draws tarot cards himself, he enters the cards into DeepSeek and asks it to interpret them. Even though different cards come up each time, DeepSeek keeps giving explanations like: 'You might have been spotted by the police, but they're too lazy to ticket you, so you get lucky.' Other times it says, 'You're fine this time.' And sometimes it says, 'This time, you're definitely getting a ticket.' It almost always matched what actually happened, so we all thought the AI was incredibly accurate.

Skepticism regarding the technology, conversely, manifests in two primary ways in Xiaohongshu. First, users frequently report specific instances of predictive failure regarding dates or past events. Comments such as, "It's inaccurate; it's already April 8th, and it said I'd receive an offer by April 7th," or "Not accurate at all; I asked it to calculate events from 20 years ago, and it claimed I was moving for school when I had actually already graduated," highlight these discrepancies. Second, and perhaps more critically, many users report that DeepSeek performs poorly at *pai pan (*the foundational step in *Bazi* divination involving the precise calculation and arrangement of the natal chart based on birth time and solar terms). As one frustrated user remarked: "This is completely useless; it arranged my *Bazi* chart entirely wrong. It might fool those who don't know better, but anyone with a bit of knowledge can see at a glance that the chart is incorrect. Even after I corrected it, the interpretation remained misleading. Promoting AI *Bazi* will really lead a lot of uninformed people astray." It is notable that such skepticism is directed towards the ability of LLMs to correctly perform an integral procedure of *Bazi* divination rather than toward the validity of *Bazi* itself, a pattern that echoes a large literature in anthropology showing that inaccurate verdicts are typically attributed to factors within the divinatory system rather than the system itself (Evans-Pritchard, 1937; Hong, 2025b; Horton, 1967).

### 4.4.2. Skepticism from human diviners: the absence of "spiritual power" and "energy"

Another major source of skepticism toward DeepSeek divination comes from human practitioners, who often critique the technology while covertly advertising their own services (as illustrated by the thematic categories in Figure 4). Like everyday users, diviners frequently point out the LLMs' struggles with accurate *pai pan* (chart calculation). However, their critiques sometimes go deeper into the ontology of divination itself. They argue that while LLMs excel at identifying and summarizing statistical regularities from internet data (and therefore mimicking entry-level diviners), these models inherently lack "spiritual power" or "energy." As one professional diviner (P2) explained:

> Even if this AI model becomes perfectly refined, it can only replace a subset of entry-level diviners. It does lean towards statistics to some extent, but as I mentioned earlier, a good fortune-teller or diviner will always possess some degree of spiritual ability. This is why divination relies on innate talent; someone might reach an entry-level proficiency [through study], but without spiritual ability, there is no way to reach the master level.

Another diviner (P5) echoed this sentiment, emphasizing that human practitioners offer an irreplaceable metaphysical connection:

> AI is a tool, not a diviner. It excels at processing structured data, but the core value of a real human divination lies in things like energy connection.

Of course, it is important to recognize that human diviners have a vested economic interest in devaluing the utility of LLM divination, and their stated opinions may serve as professional boundary-work rather than purely objective critiques. Even so, the recurring insistence on non-material "talent" and spiritual resonance is not limited to professionals. It reflects a broader cultural schema that also appears throughout our interview data. As one interviewee (U4) put it:

> So divination itself, fortune-telling itself, these numerology techniques, are capacities that belong to humans. And for things like *Bazi* divination, there's *energy*; a person's energy is very important. For example, some practitioners of Chinese medicine also combine *Bazi* when they assess someone, there's observation, listening/smelling, inquiry, and palpation. That 'observation' is reading with energy, not with the naked eye. It's like opening a 'third eye' or something: you can perceive an energy field. Some people really can see it directly—see a person's meridians, see this energy field, see the aura of the body, see what someone is emitting. But AI definitely can't.

This account is revealing in two ways. First, it articulates a theory of efficacy grounded in embodied perception and energetic attunement (e.g., "energy fields," "aura," "third eye"), treating divinatory accuracy as inseparable from the practitioner's cultivated capacities. Second, it draws a sharp boundary between human and machine by locating the decisive ingredient of divination in a non-computational domain: sensory–spiritual access to otherwise invisible realities. Within this framework, even if an LLM could perfectly reproduce the textual conventions of Xuanxue discourse, it would still be disqualified as a “true” diviner because it cannot participate in the metaphysical relations presumed to underwrite divinatory knowledge.

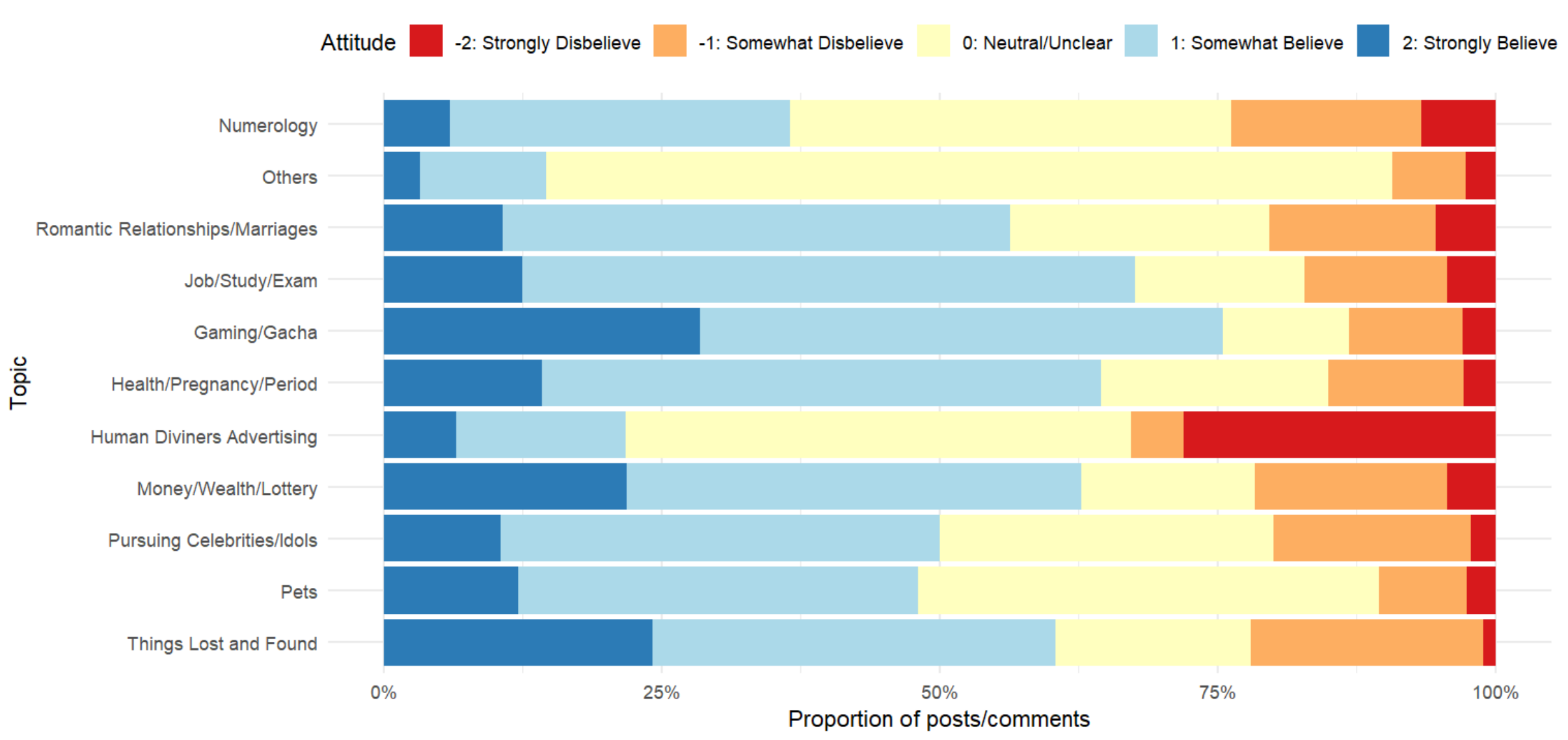

*Figure 4. Attitudes toward the efficacy of DeepSeek Xuanxue in Xiaohongshu by topic. Colors indicate stance scores on a 5-point scale, and segment widths represent the proportion of posts/comments within each thematic topic category.*

### 4.4.3. Repeatability, consistency, and cross-Model comparison

A distinctive affordance of LLM-based *Xuanxue* is its virtually zero marginal cost for producing additional readings. Because users can ask the same question repeatedly at no extra expense (especially given that DeepSeek remains free for most users in China, unlike consultations with human practitioners), they often “test” the system by submitting identical queries across different chat sessions. This practice echoes classic divinatory logics in which repeated trials and convergence across consultations function as a basis for credibility (Hong & Henrich, 2021, 2025; Nadel, 1954; Raphals, 2013). In this context, output consistency becomes a primary heuristic for trustworthiness. When the model fails to produce stable answers, users quickly voice their skepticism. For instance, one user noted, “*I put in the Four Pillars (year, month, day, hour), asked the same question in two different conversations, and got completely different answers.*” Another shared a similar disillusionment: “*The first time I calculated it, the result was pretty close to what a human diviner told me. But the second time I asked in the exact same way, the result was completely the opposite.*” Others warned their peers about this algorithmic instability, with comments such as “*Not accurate, three tries, three different answers*” and “*Accurate? I sent the same instruction twice and got different answers*”.

Verification practices also extend beyond repeated sessions on a single platform to cross-model comparison. As alternative LLMs became more accessible in China in the latter half of 2025 (often via VPNs or proxy accounts), users increasingly triangulated their readings across systems such as DeepSeek, *Doubao*, Gemini, and ChatGPT. For some, convergence across models is treated as evidence of validity. One user wrote: "*Lately I've been obsessed with using AI to predict career luck and wealth luck, DS, Doubao, and Gemini are all roughly consistent on the big picture, so it should be accurate*". For others, divergence across models serve as evidence against accuracy: "*Same birth chart, two different AIs, completely different results. [Therefore, don't believe it.]*"

In the supplementary dataset, we observed this comparative stance becoming more explicit, and discussions often focused on the relative capability and accuracy of different models. Under a post titled "*I think Gemini is currently the most accurate,*" numerous commenters spontaneously debated the divinatory performance of DeepSeek versus Gemini. One user wrote, "*Gemini seems overly reliant on the balance of vitality and decline. Its judgments about having a 'strong constitution' or 'weak constitution' don't fit me, DeepSeek is more accurate.*" Another described a more iterative, adversarial form of testing: "*I pasted into Gemini the exact wording DS used, and Gemini refuted everything DS said. Then I fed Gemini's rebuttal back to DS, and DS admitted it hadn't considered this or that, and ultimately affirmed Gemini's view. So it seems Gemini has a slightly broader knowledge base than DS.*" A third user offered a contrasting experience regarding technical precision: "*My birthday falls right on the cusp of a solar term. Gemini messed up the chart completely and refused to correct it, so the results were almost entirely borderline or wrong. Only GPT worked well.*" These exchanges illustrate how users evaluate LLM-based divination not only through the felt resonance of a single reading but also through rigorous comparative prompting across systems.

Consequently, many threads exhibit a pronounced *trial-and-error* ethos, as users test models, swap prompts, and troubleshoot failures collectively. This collaborative debugging highlights the instrumental character of contemporary LLM divination: users engage these systems as configurable tools whose outputs can be improved (at least in principle) through the right platform choice and prompting strategy. For example, under a post where a user lamented, "*I sent the exact same Bazi twice and got different results… now I'm hesitating,*" another replied pragmatically: "*Try it again with GPT.*" Across such interactions, "efficacy" is negotiated through iterative experimentation and comparison. Replication and cross-platform triangulation thus operate as everyday verification practices, through which users assess both divinatory claims and the models themselves by checking whether results converge across sessions and platforms.

### 4.4.4. The tension between *Xuanxue* and Science

Lastly, we examine the larger epistemological frameworks that users apply in understanding LLM outputs, as the perceived legitimacy of LLM-based *Xuanxue* (or any other divinatory method; see Hong & Henrich, 2021) depends not only on empirical "accuracy" or cross-platform consistency, but also on users' underlying worldviews, especially their assumptions about what kinds of causal relations are possible in the world. On Xiaohongshu, users frequently highlight the technical nature of their practices through hashtags. They routinely combine model names like #DeepSeek or #ChatGPT with terms such as #AITools, #Algorithm, and #Cyber. By doing so, they rhetorically distance their activities from traditional superstition, framing them as computationally mediated and inherently systematic. The use of tags like #ScientificPregnancyPlanning on divination posts, for example, serves to reposition ritualistic prediction as a form of technical planning. As Gillespie (2014) notes, invoking the term "algorithmic" often functions as a credibility cue, lending an aura of objectivity and trustworthiness to the output even when the underlying mechanism is opaque.

Yet this appeal to science often coexists with (and sometimes even gives way to) a counter-frame in which metaphysics is treated as *more* powerful than scientific reasoning. A notable subset of posts utilizes hashtags such as "*The end point of science is Xuanxue*" or "*The ultimate truth of the world is Xuanxue.*" Rather than borrowing scientific authority, these tags reverse the hierarchy, asserting that *Xuanxue* offers an explanatory power that supersedes scientific reasoning. Importantly, the "science" being contrasted here is typically understood as *current* science (i.e., a necessarily incomplete account of the material world), so "*Xuanxue*" becomes a catch-all for what has not (yet) been explained within that framework.

Beyond slogans, users also discuss LLM divination in explicitly "big data" terms, framing both its promise and its limits. Some commenters treat divination as fundamentally compatible with computation and expect future models to approximate most practitioners, though not exceptional "gifted" masters:

> As far as I understand it, fortune-telling is basically math, formulas and parameters. So maybe DeepSeek has only learned a small part of those formulas so far. If it learns the full set later, it should be able to match most masters, though probably not the very top, naturally gifted ones.

Another commenter offered a more pragmatic, tool-like view that treats the model as technically useful but in need of correction: "*AI is a big-data model after all; sometimes you have to correct it.*" Other users, however, mobilize the same "big data" explanation to *dismiss* LLM divination as mere pattern-matching dressed up in metaphysical language. For instance, one comment reduces the practice to analytics "under the banner" of *Bazi*: "*It's just big-data analysis using Bazi as a pretext*". A similar critique frames any publicly learnable "algorithm" as inherently shallow, implying that real expertise cannot be compressed into something a model can reproduce for everyone: "*AI's fate-calculation 'algorithms' come from integrating and processing big data. If it's something ordinary people can learn from data, it's only the surface level.*" Finally, a third strand of comments pushes back against the big-data account by pointing to experiences of uncanny personal accuracy. Here, the surprising specificity of a reading is taken as evidence that something more than generic statistical inference might be at work:

> Mine was incredibly accurate… It even knew I transferred schools in fourth grade. I'm starting to wonder whether it really is 'just' big data.

These discussions show that users of DeepSeek *Xuanxue* recognize yet show limited appreciation of the data-driven processing of LLMs in generating divinatory verdicts. More generally, "science" in DeepSeek *Xuanxue* discourse rarely signals a firm commitment to a fully mechanistic worldview, but more often functions as a flexible label for interpretation and legitimation of seemingly irrational practices. While the friction between *Xuanxue* and science occasionally surfaces as an ideological debate (such as when users directly ask "*So tell me, is this science or Xuanxue?*" or "*If it (LLM divination) reaches this level [of technical sophistication], then fortune-telling is no longer Xuanxue, but rather science.*"), participants are generally comfortable tolerating and actively leveraging a significant degree of epistemological ambiguity, exemplified by posts like "*Science and Xuanxue were never opposites to begin with.*"

# 5. Discussion

## 5.1. Divination in the age of Large Language Models: uncertainty and instrumentality

Given the near-ubiquitous presence of divination across human societies (Hong, 2025a), it is perhaps unsurprising that *Xuanxue*—specifically divination—has emerged as one of the most prominent use cases for generative AI among Chinese youth. A rich body of literature in the cognitive science of religion highlights the psychological mechanisms that render such divinatory practices intuitively plausible (Boyer, 2020; Hong & Henrich, 2021). The rise of AI-mediated divination adds important nuances to grand theories of secularization (Bruce, 2011), contributing to a growing body of scholarship on the persistence and adaptation of spiritual practices within technologically sophisticated societies (e.g., Campbell & Tsuria, 2021; McClure, 2017)

We would like to emphasize that, despite its obvious entertainment value and playful platform dynamics, LLM divination exhibits many of the core functional features of traditional divinatory practices. Users turn to these models to navigate domains where empirical information is scarce or unavailable, seeking guidance on everything from romantic entanglements and career prospects to the optimal timing for in-game "gacha" draws. Underlying this turn to divination is a clear psychological need; as Pan et al. (2025) suggest, the contemporary *Xuanxue* trend reflects an attempt to alleviate anxiety and obtain a sense of (albeit sometimes illusory) control in an increasingly precarious socioeconomic environment. Indeed, our interviewees frequently and explicitly cited situational anxiety as their primary motivation for engaging with LLM divination, leveraging it as a convenient, zero-cost mechanism for managing stress.

However, it is important to note that a substantial proportion of both Xiaohongshu users and our interviewees genuinely believe in the validity of the divinatory outputs, granting the AI-generated information real epistemic value. Such attitudes align with a robust anthropological literature on the intellectual interpretation of magico-religious systems (Hong et al., 2024; Hong & Henrich, 2025; Horton, 1967; Jarvie, 2018), which posits that practitioners engage with these systems not merely for emotional comfort, but because they perceive them as instrumentally effective tools for understanding and navigating the world. In fact, as we have pointed out emphatically, functional explanations make sense only in light of the actors' (at least partial) belief in the efficacy of these practices. This is because the functional consequences of divination practices crucially depend on people's belief in such practices fulfilling their instrumental goals. In the case of divination, this means that their psychological and social functions depend on people placing some faith in divination's epistemic value (Hong, 2025a).

## 5.2. The perception of predictive accuracy: Barnum effect, selective reporting, and selection biases

As detailed in our results, a substantial portion of users who express a clear stance on the efficacy of LLM divination report that DeepSeek is highly (sometimes "scarily") accurate. To understand why users so frequently report these "accurate predictions" given the scientific impossibility of divination, we must look beyond the computational capabilities of the models and examine the psychological and sociological dynamics at play, compounded by platform-specific factors.

Psychologically, the widespread perception of accuracy can be largely attributed to a combination of the Barnum (or Forer) effect and confirmation bias (Forer, 1949; Meehl, 1956). By design, Large Language Models are engineered to generate plausible and contextually adaptable text. When prompted for a divinatory reading, the AI often produces outputs that contain generalized, universally applicable statements alongside probabilistic guesses. Users then actively participate in the meaning-making process by mapping these outputs onto their own highly specific, personal situations, a classic manifestation of the Barnum effect. Furthermore, users engage in retrospective confirmation: they selectively remember and highlight the specific details that align with their reality (the "hits," such as the model seemingly knowing about a childhood school transfer) while rationalizing or simply forgetting the inaccuracies (the "misses") (Hong, 2022, 2024).

Sociologically, similar to many other technological practices (de Barra, 2017; Hong, 2022; Hong & Zinin, 2023), the dynamics of social media platforms heavily incentivize selective reporting, which artificially amplifies the perceived efficacy of LLM divination in the public sphere. Users are more motivated to draft a post or leave a comment about an astonishingly accurate prediction than to document a mundane, generic, or incorrect one. Consequently, the public discourse surrounding DeepSeek *Xuanxue* becomes saturated with success stories, creating a social proof feedback loop that encourages new users to approach the AI with heightened expectations of accuracy. Importantly, this is not to say that inaccurate predictions go entirely unreported (as shown in Section 4.4). Rather, when predictive failures do occur, they are frequently rationalized through "auxiliary hypotheses" (e.g., incorrect prompts). As is common in many religious and ritual contexts, these rationalizations protect the fundamental validity of the divinatory system itself from being questioned (Hong, 2025b).

Finally, it is important to interpret our findings through the lens of selection bias. The users actively participating in these Xiaohongshu exchanges form a highly self-selecting group; people who are secular, uninterested in Xuanxue, or already skeptical of LLM divination are simply unlikely to engage with or comment on these posts in the first place. Because our corpus naturally over-represents enthusiasts and those already open to mystical frameworks, the high prevalence of belief observed in our data (as shown in Figure 3) must be carefully contextualized. As such, caution should be exercised when attempting to extrapolate these findings to the broader population in China or elsewhere. Furthermore, Xiaohongshu's recommendation algorithms are designed to maximize engagement, meaning that sensational and emotionally resonant posts are systematically pushed to the top of users' feeds.

## 5.3. *Xuanxue*, science, the possibility of LLM divination and its societal impact

In one sense, the use of LLMs for divination is unsurprising. Across history and societies, people have sought answers under conditions of uncertainty, especially when the perceived stakes are high (Hong, 2025a). LLMs, precisely because they are flexible and general-purpose, can readily be recruited as sources of actionable guidance, even when that guidance is not scientifically warranted. Their apparent predictive power is further reinforced by the psychosocial dynamics discussed in the previous section, as well as by the longstanding intuitive appeal of established divinatory traditions. In our materials, this includes Chinese practices such as *Bazi* numerology (which draws much of its authority from the *Yijing*) and Western esoteric practices such as tarot and astrology. As shown in Section 4.4, while many users describe LLM *Xuanxue* as "mechanical" and therefore lacking the "spiritual energy" attributed to skilled human diviners, they nonetheless often judge it to be "good enough" for everyday decision-making.

The fact that some participants perceive a tension between *Xuanxue* and science is also noteworthy. On the one hand, hashtags such as "*The end point of science is Xuanxue*" explicitly frame the two as oppositional, casting *Xuanxue* as addressing what science cannot (yet) explain. On the other hand, posts and interview responses suggest that many users are less invested in adjudicating deep ontological disagreements than in using "science vs. *Xuanxue*" as a rhetorical frame that can be invoked or downplayed depending on context. This pattern aligns with findings in cognitive psychology showing that people can hold natural and supernatural explanations in parallel, drawing on each in different situations and using a range of strategies to manage potential conflicts between them (Legare et al., 2012; Legare & Gelman, 2008).

It remains to be seen how the availability of LLMs affects the public perception of both specific divinatory practices and *Xuanxue* in general. At the government level, the regulation of such potentially "superstitious" content on public platforms has always existed (G. Li, 2018), but how LLMs should generate *Xuanxue* content is currently somewhat of a grey area. As a result, restrictions in this area are largely implemented through platforms' own compliance and content-moderation choices during the operation of domestic LLM and generative AI services (e.g., DeepSeek). This compliance posture is shaped in part by the *Interim Measures for the Management of Generative AI Services* issued by the Cyberspace Administration of China (2023), which requires generative AI providers to ensure that generated content is "positive, healthy, and uplifting" and is subject to appropriate content-governance mechanisms. Against this backdrop, esoteric/occult content may be treated by platforms as a compliance risk and, consequently, is sometimes automatically filtered or constrained by the model's safety and moderation systems.

In practice, however, many users can still pose (and obtain responses to) *Xuanxue* questions with sufficiently careful prompting. LLM divination may therefore have meaningful cognitive and societal effects as it changes the conditions under which divinatory guidance and verdicts are accessed and evaluated. First, it reduces friction: readings become inexpensive and available on demand, and can be repeated at will, which can increase the frequency with which users consult divination for everyday uncertainty. Second, it changes the *interactional format* of divination. Unlike many offline consultations, LLM divination is easily iterated: users can users can refine prompts and request alternative interpretations until an output "fits", a dynamic that may amplify confirmation bias and hindsight rationalization. Third, because the model can produce fluent, personalized narratives at scale, it may strengthen the subjective sense of being "seen" or accurately diagnosed, even when the inputs are minimal—an effect that can deepen perceived efficacy and encourage reliance. In benign cases, this serves as low-stakes emotional support or a way of organizing reflection. In more problematic cases, it may promote overconfidence in unverifiable claims, normalize the outsourcing of consequential decisions (e.g., those regarding relationships, education, or employment) to automated readings, or intensify anxiety through repetitive checking and rumination.

At a societal level, the same features that make LLM divination attractive (scalability and low cost) also make it capable of reshaping how spiritual and quasi-spiritual practices circulate online (Campbell & Tsuria, 2021). When packaged in the idiom of data, algorithms, and "AI," divinatory output can be presented as a modern, technical service rather than as superstition; yet when its claims are questioned, it can just as readily be reframed as entertainment or "only for reflection," avoiding direct accountability. This combination of scientific-sounding legitimation and easy retreat into disclaimers can blur the boundaries between entertainment, self-help, and truth claims, making it harder for audiences (and platforms) to determine whether a reading is being treated as playful content or as guidance carrying genuine epistemic authority. Finally, because content moderation can unevenly constrain what can be asked and answered, platform governance may indirectly steer which forms of *Xuanxue* appear legitimate, which styles of explanation proliferate, and what kinds of claims users learn to present as acceptable, thereby shaping not only *whether* LLM divination spreads, but also *what kind* of divination it becomes.

# 6. Conclusion

This study has offered one of the first systematic examinations of LLM-based divination, drawing on a corpus of Xiaohongshu posts and semi-structured interviews to document how Chinese youth engage with DeepSeek as a divinatory tool.

Our analysis shows that LLM divination, despite its novel technological substrate, reproduces the core functional logic of traditional divinatory practices: users consult it under conditions of uncertainty, grant its outputs genuine epistemic value, and evaluate its efficacy in instrumental terms. This continuity suggests that divination is best understood not as a relic of pre-modern thought but as a durable cognitive-instrumental strategy that adapts fluidly to new mediating technologies. At the same time, the phenomenon reveals how users appropriate general-purpose conversational AI for purposes far beyond its intended scope, repurposing LLMs as divinatory agents through creative prompt engineering. The persistence of this perceived accuracy has less to do with any real predictive power on the part of LLMs than with a set of reinforcing psychological, sociological, and platform-specific mechanisms (the Barnum effect, confirmation bias, selective reporting, and selection bias) that together sustain and amplify the sense that AI-generated readings carry genuine epistemic weight. These findings bridge the anthropology of divination and the study of human-AI relations by showing that users ascribe quasi-agentive authority to AI outputs through the same interpretive and evaluative practices documented in ethnographies of traditional divination. The case of LLM divination thus illuminates both the persistence of divinatory reasoning in contemporary life and the culturally embedded ways in which people negotiate meaning, trust, and agency in their encounters with AI.

Several limitations point toward productive avenues for future research. Our data capture a single, rapidly evolving moment in the adoption of generative AI; longitudinal work is needed to determine whether LLM divination stabilizes as a durable practice or proves to be a transient novelty. Cross-platform and cross-cultural comparisons would help clarify the extent to which the dynamics we observe are specific to Xiaohongshu's affordances and to the Chinese *Xuanxue* tradition, as opposed to reflecting more general tendencies in how people relate to AI. Experimental approaches could isolate the causal contribution of the Barnum effect and confirmation bias with greater precision than observational data allow. Finally, future work should examine the downstream behavioral consequences of LLM divination, including its effects on decision quality, anxiety regulation, and broader patterns of AI reliance.

## Data availability

All data and analysis files are available at https://osf.io/x47ey/overview?view_only=ce283d06e31249119df7ea1378e1f034

## Declaration of generative AI and AI-assisted technologies in the manuscript preparation process

During the preparation of this work the author(s) used Gemini-3.0-Pro and Claude-Opus-4.6 in order to translate the interview protocol and proofread the manuscript. After using this tool/service, the author(s) reviewed and edited the content as needed and take(s) full responsibility for the content of the published article.